\documentclass[lettersize,journal,onecolumn]{IEEEtran}
\usepackage{amsmath}
\usepackage{amsfonts}
\usepackage{subfig}
\usepackage{amssymb}
\usepackage{multirow}
\usepackage{color}
\usepackage{amsthm}
\usepackage{hyperref}
\usepackage{algorithmic}
\usepackage{algorithm}
\usepackage{array}
\usepackage{stfloats}
\usepackage{url}
\usepackage{verbatim}
\usepackage{caption}
\usepackage{graphicx}
\usepackage{xcolor}
\usepackage{cite}
\usepackage{flushend}
\usepackage{soul}
\usepackage[normalem]{ulem}
\newcommand{\stkout}[1]{\ifmmode\text{\sout{\ensuremath{#1}}}\else\sout{#1}\fi}
\begin{document}

\title{Handling~Interference~in~Integrated~HAPS-Terrestrial Networks through Radio Resource Management}
\author{Afsoon Alidadi Shamsabadi,~\IEEEmembership{Member,~IEEE,}~Animesh~Yadav,~\IEEEmembership{Senior~Member,~IEEE,}~Omid~Abbasi,~\IEEEmembership{Student Member,~IEEE,} Halim Yanikomeroglu,~\IEEEmembership{Fellow,~IEEE}
\thanks{This work has been supported by Huawei Canada Co. Ltd.

All the authors are with the Department of Systems and Computer Engineering, Carleton University, Ottawa, ON, K1S 5B6, Canada. (e-mail: \{afsoonalidadishamsa, animeshyadav, omidabbasi, halim\}@sce.carleton.ca).}
}
\maketitle

\begin{abstract}
Vertical heterogeneous networks (vHetNets) are promising architectures to bring significant advantages for 6G and beyond mobile communications. High altitude platform station (HAPS), one of the nodes in the vHetNets, can be considered as a complementary platform for terrestrial networks to meet the ever-increasing dynamic capacity demand and provide sustainable wireless networks for future. However, the problem of interference is the bottleneck for the optimal operation of such an integrated network. Thus, designing efficient interference management techniques is inevitable. In this work, we aim to design a joint power-subcarrier allocation scheme in order to achieve fairness for all users. We formulate the max-min fairness (MMF) optimization problem and develop a rapid converging iterative algorithm to solve it. Numerical results validate the superiority of the proposed algorithm and show better performance over other conventional network scenarios.
\end{abstract}


\section{Introduction}
The demand for ubiquitous connectivity, the support for dynamic and unpredictable capacity demands, and the emergence of novel use cases will necessitate the integration of various network tiers for 6G and beyond mobile communications. For example, heterogeneous networks (HetNets) integrate terrestrial macro and small cell tiers, and enable dense cellular network to improve the coverage and spectral-efficiency (SE) for edge users\cite{ref1}. However, terrestrial network densification alone cannot meet network's sustainability requirement, cater capacity demand for ever-increasing users, and support novel use cases. Therefore, integration of terrestrial network with a complementary platform, which can deploy reusable energy sources to support sustainability and complement terrestrial network to offer larger capacity and coverage, is an obvious solution\cite{ref2}.

High altitude platform station (HAPS) is a quasi-stationary node in the stratosphere layer at an altitude 20 km above the ground and offers many advantages, such as i) it has lower latency than artificial satellites; ii) it offers flexibility in supporting sophisticated technologies (e.g., massive multiple-input multiple-output (MIMO) and reconfigurable intelligent surfaces, etc.); iii) it has larger coverage area than terrestrial base stations. Therefore, HAPS can be considered as a complementary platform that has the potential to meet the future network requirements together with legacy terrestrial and space networks\cite{ref3}. Furthermore, the large surface area of HAPS facilitates the deployment of ambient energy harvesting resources, which helps in creating a sustainable network architecture.

Integration of HAPS with existing multi-tier networks has been considered a promising architecture for future networks \cite{Covextension, NOMA}. The authors in \cite{Covextension} proposed a cooperative non-terrestrial and terrestrial network focusing on HAPS use cases and technical issues. The authors in \cite{NOMA} considered an integrated satellite-HAPS-terrestrial network architecture and investigated the performance of transmit antenna selection schemes in non-orthogonal multiple access (NOMA) scenarios.

However, multi-tier networks face various types of challenges and interference is the most serious one. In this context, efficient interference mitigation mechanisms, such as user association, radio resource management and interference alignment, should be designed and utilized in the networks\cite{ref4}. Diverse interference management techniques have been proposed for HetNets which require coordination among network tiers to optimally design the transmission parameters and resources\cite{ref5,ref1}. For vertical heterogeneous networks (vHetNets), the challenge of interference gets more complicated and challenging, and urges for more novel techniques to deal with it. To this end, the authors in\cite{ref6} proposed a multi-agent deep Q-learning (DQL)-based transmit power control algorithm to minimize the outage probability of the HAPS downlink channel under the constraint of interference tolerance of the terrestrial network. The authors in \cite{ref7} designed transmit beamformers to mitigate interference from the HAPS to terrestrial cells. The beamformers form nulls toward terrestrial cells, while forming user-specific beams towards the user equipment (UEs) associated with the HAPS. The authors in \cite{ref7a} considered the coexistence of HAPS and terrestrial networks and analysed the impact of interference in the coexisting network. Their results suggested some minimum inter-site distance that should be maintained between HAPS and terrestrial network to mitigate interference. The authors in \cite{ref7b} proposed an interference coordination method for an integrated HAPS-terrestrial network considering the distribution of traffic load between HAPS and terrestrial network.

In this work, we are interested in mitigating interference in a vHetNet, consisting of a HAPS and a few terrestrial macro base stations (MBSs). However, the different transmit characteristics, such as transmit power and antenna gain, of MBSs and HAPS lead to unfairness among UEs. Therefore, we use max-min fairness (MMF) as the objective function of our optimization problem to improve fairness. 
The main contributions in this paper include the following:
	\begin{itemize}
		\item We formulate a novel optimization problem to design the joint subcarrier and power allocation in vHetNets by maximizing the minimum achievable rate of UEs.
		\item Since the formulated problem is mixed-integer non-linear program (MINLP), we develop a low-complexity and rapid converging iterative algorithm based on successive convex approximation (SCA) framework.
		\item We provide comprehensive simulation results to compare the performance of the proposed algorithm in vHetNet with conventional schemes and system architectures. The results show the performance improvement, achieved by the proposed algorithm in vHetNet.
	\end{itemize}
	
The remainder of the paper is organized as follows.  Section \ref{Sec:model} presents the system model and formulates the optimization problem. Section \ref{Sec:Problem_sol} details the proposed solution and algorithm. Section \ref{Sec:Results} presents and discusses the obtained numerical results. Finally, Section \ref{Sec:conclution}  concludes the paper.

\section{System Model and Problem Formulation}\label{Sec:model}
\begin{figure*}[t]
\centering
\captionsetup{justification=centering}
\includegraphics[width=0.5\linewidth]{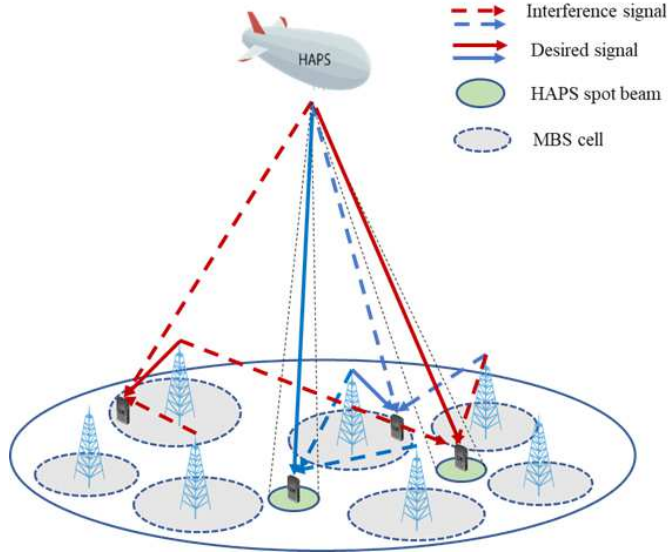}
\caption{\small System model of an integrated network (vHetNet).}
\label{fig_1}
\end{figure*}

Consider a vHetNet consists of one multi-antenna HAPS and $N_\text{B}$ multi-antenna MBSs. The HAPS and MBSs simultaneously serve $N_\text{u}$ single antenna UEs in an overlapped geographical area, as depicted in Fig.~\ref{fig_1}. The base stations and UEs are indexed as $j\in\{1,\ldots, N_{\text{B}}+1\}$ and $i\in\{1,\ldots, N_{\text{u}}\}$, respectively. Base station index $j=N_{\text{B}}+1$ is reserved for the HAPS. The vHetNet employs orthogonal frequency division multiplexing (OFDM) transmission scheme with a total of $N_{\text{f}}$ subcarriers, which are shared among the HAPS and MBSs. The subcarriers are indexed as $f\in\{1,\ldots, N_{\text{f}}\}$. Further, we assume that the vHetNet operates in the downlink subframe.

The multi-antenna MBS is equipped with $N_{\text{R}}$ elements, arranged in a uniform linear array (ULA) configuration. The $N_{\text{R}}$ elements are used for beamforming towards different UEs to manage interference more efficiently. The MBSs apply maximum ratio combining (MRC) precoding scheme before transmitting the data.
We consider free-space path loss (FSPL) and slow Rayleigh fading channel between the UEs and MBSs. We denote $h_{r,i,j,f}\sim \mathcal{NC}(0,1)$ as the Rayleigh fading channel gain between UE $i$ and antenna element $r$ of MBS $j$  on subcarrier $f$.

On the other hand, the multi-antenna HAPS is equipped with $N_{\text{V}}$ and $N_{\text{H}}$ elements in the vertical and horizontal axis, respectively, which are arranged in a uniform planar array (UPA) configuration. Following the ITU technical report \cite{ref8}, the UPA configuration is able to generate the antenna pattern $G(\theta,\phi)$ to form a spot beam\footnote{Spot beam is the steered power concentrated signal covering a limited geographical area.} toward a UE on the ground. We denote the antenna pattern that is used to form the spot beam towards UE $i$ by $G_i(\theta,\phi)$ with the maximum gain at $(\theta_i,\phi_i)$, which are the elevation and azimuth angle pair of UE $i$. The HAPS spot beams are generated on the basis of capacity demand of an associated UE and can overlap with each other. 
Because of the HAPS altitude, all the UEs form LoS channel with the HAPS and FSPL is the dominant loss of this channel.

Since the HAPS and MBSs are simultaneously serving the grounds UEs, the UEs can get service from either of them that satisfies their rate requirement. For example, a UE will be served by the HAPS if there is not enough coverage or capacity from the MBSs, otherwise by the MBSs. To serve an UE, the HAPS generates dynamic spot beam toward it.

In this work, we allocate orthogonal subcarriers to UEs within the MBS cell or HAPS coverage area and each UE is supposed to be allocated only one subcarrier. Consequently, there is no intra-cell interference in terrestrial network cell and HAPS coverage area. Further, the HAPS and terrestrial network share the same set of subcarriers; therefore, the UEs associated with the HAPS will receive interference from nearby MBSs. Similarly, the UEs associated with the MBSs receive interference from the HAPS and other MBSs. It is worth noting that, in this case, the received interference from the HAPS is much stronger due to LoS connection and higher antenna pattern gain. Based on the previous discussion, the SINR of UE $i$ can be expressed as
\begin{equation}\label{eq:SINR}
\gamma_i=\cfrac{P_{\text{R}_i}}{I_i+P_{\text{N}}},
\end{equation}
where $P_{\text{R}_i}$ and $\text{I}_i$ denote the desired and interference signal power of UE $i$, respectively. $P_{\text{N}}$ is the noise power over one subcarrier. Specifically, $P_{R_i}$ and $I_i$, can be expressed, respectively, as
\begin{equation}{\label{eq5}}
P_{\text{R}_i}=(1-A_{i,{N_{\text{B}}+1}})\sum _{f=1}^{N_{\text{f}}}{\frac{F_{i,f}P_i\sum _{r=1}^{N_{\text{R}}}{\left|h_{r,i,\text{J}_i,f}\right|^2}}{\text{PL}_{i,\text{J}_i}}}
+A_{i,N_{\text{B}}+1}\frac{P_iG_i(\theta _i,\phi _i)}{\text{PL}_{i,N_{\text{B}}+1}},
\end{equation}
\begin{equation}{\label{eq6}}
I_i=\sum_{k=1,i\not=k}^{N_{\text{u}}}(1-A_{k,N_{\text{B}}+1})
\frac{\sum_{f=1}^{N_{\text{f}}}{F_{i,f}F_{k,f}P_k\sum_{r=1}^{N_{\text{R}}}{\left|h_{r,i,J_k,f}h^*_{r,k,J_k,f}\right|}}}{\text{PL}_{i,J_k}}
+A_{k,N_{\text{B}}+1}\frac{G_k(\theta _i,\phi _i)\sum_{f=1}^{N_{\text{f}}}{F_{i,f}F_{k,f}P_k}}{\text{PL}_{i,J_k}},
\end{equation}
where $P_i$ denotes the transmit power allocated to UE $i$, and $\text{PL}_{i,j}$ denotes the FSPL between user $i$ and base station $j$, and is expressed as 
\begin{equation}{\label{eq1}}
   \text{PL}_{i,j}=(\frac{4\pi f_c d_{i,j}}{c})^2,
\end{equation}
where $f_c$ is the carrier frequency (in Hz), $d_{i,j}$ (in m) is the distance between the UE $i$ and base station $j$, and $c$ represents the speed of light in free space. $F_{i,f}$ denotes the indicator variable for subcarrier allocation such that 
\begin{equation}{\label{eq:subcarrier_indicator}}
{F_{i,f}} = \begin{cases}
{1,}&{\text{if the $f$th subcarrier is allocated to UE $i$}},\\ 
{0,}&{\text{otherwise.}}
\end{cases}
\end{equation}
Further, $A_{i,j}$ is the element of the UE association matrix, $\mathbf{A} \in \mathbb{B}^{N_{\text{u}}\times ({N_{\text{B}}+1})}$, that maps the base stations with the UEs. Particularly, if UE $i$ is associated with base station $j$, then $A_{i,j}$ is $1$, otherwise 0. ${J}_i\in \{1, \ldots, N_{\text{B}}+1\}$ is the index of the base station to which UE $i$ is associated.
The matrix $\mathbf{A}$ is determined based on maximum received SINR, from all the base stations including HAPS, considering maximum load of each base station, which is equal to $N_{\text{f}}$. 

Now, we formulate the MMF optimization problem that designs power and subcarrier allocation in every subframe. Specifically,  we maximize the minimum achievable rate of the network, which is equivalent to maximizing the minimum SINR of the UEs. The optimization problem can be expressed as
\begin{IEEEeqnarray*}{lcl}\label{eq:P1}
&\underset{\mathbf{F},~\mathbf{P}}{\text{maximize}}\,\,~ & \min_{\forall i}\quad \gamma_i  \IEEEyesnumber \IEEEyessubnumber* \label{eq:P1_Obj}\\
&\text{s.t.} & N_{\text{R}}\sum_{i=1}^{N_{\text{u}}}{A_{i,j}P_i}\leq P_\text{t}^{\text{MBS}}, \, ~\forall j\in\{1,\ldots, N_{\text{B}}\},\label{eq:P1_const1}\\
&& \sum_{i=1}^{N_{\text{u}}}{A_{i,N_{\text{B}}+1}P_i}\leq P_\text{t}^{\text{HAPS}}, \,\label{eq:P1_const2}\\
&& \sum_{j=1}^{N_{\text{B}}+1}{\sum_{k=1,k\not=i}^{N_{\text{u}}}{A_{i,j}A_{k,j}\sum_{f=1}^{N_{\text{f}}}{F_{i,f}F_{k,f}}}}=0, \, ~\forall i, \qquad \label{eq:P1_const3}\\
&& \sum_{f=1}^{N_{\text{f}}} F_{i,f}=N_{\text{SC}}, \, ~\forall i, \label{eq:P1_const4}\\
&& F_{i,f} \in \mathbb{B}, \, ~\forall i, ~\forall f, \label{eq:P1_const5}\\
&& P_i \geq 0, \, ~\forall i, \label{eq:P1_const6}
\end{IEEEeqnarray*}
where $\mathbf{F} \in \mathbb{B}^{N_{\text{u}} \times N_{\text{f}}}$ denotes the binary subcarrier allocation matrix and its elements are defined according to \eqref{eq:subcarrier_indicator}. $\mathbf{P}= [P_1, \ldots,P_{N_{\text{u}}}]^T \in \mathbb{R}^{N_{\text{u}}\times 1} $ denotes the UE power allocation vector. $P_\text{t}^{\text{MBS}}$ and $P_\text{t}^{\text{HAPS}}$ are the maximum total transmit power of the MBS and HAPS, respectively. $N_{\text{SC}}$ refers to the number of subcarriers, allocated to each UE.

In problem \eqref{eq:P1}, constraints (\ref{eq:P1_const1}) and (\ref{eq:P1_const2}) limit the maximum transmit power of the MBS and HAPS. Constraint (\ref{eq:P1_const3}) ensures that orthogonal subcarrier allocation has been preserved within each MBS and HAPS. Constraint (\ref{eq:P1_const4}) restricts each UE to use only $N_{\text{SC}}$ subcarriers. Constraints (\ref{eq:P1_const5}) and (\ref{eq:P1_const6}) ensure that each element of matrix $\mathbf{F}$ takes only binary values and $\mathbf{P}$ values are positive, respectively. 

Due to the presence of binary variable $\mathbf{F}$, and non-convex objective function and constraint \eqref{eq:P1_const3}, problem \eqref{eq:P1} is a mixed-integer non-linear program (MINLP), which is challenging to solve for the global solution. Thus, in the following section, we develop a SCA-based rapid converging iterative algorithm to solve \eqref{eq:P1} for the sub-optimal solution. To this end, we apply the reformulation-linearization techniques (RLT) to transform the MINLP into an equivalent convex problem, which is then solved in an iterative manner until it converges to the sub-optimal solution of problem \eqref{eq:P1}.
\vspace{-0.17cm}
\section{Proposed Solution}\label{Sec:Problem_sol}
In this section, we equivalently transform problem (\ref{eq:P1}) into the mathematically tractable form and apply a few approximations to deal with non-convex parts.
First, by introducing two slack variables $z_i,~\forall i,$ and $\beta_i,~\forall i,$ we equivalently replace the objective function \eqref{eq:P1_Obj} with $z_i$ along with  two new constraints as follows:
\begin{IEEEeqnarray*}{lcl}\label{eq:obj_transform}
\beta_i z_i\leq P_{R_i}, \IEEEyesnumber \IEEEyessubnumber* \label{eq:obj_transform_a}\\
\beta_i\geq I_i+P_{\text{N}}. \label{eq:obj_transform_b}
\end{IEEEeqnarray*}

Note that the objective function is convex now, however, \eqref{eq:obj_transform_a} and \eqref{eq:obj_transform_b} are both non-convex constraints.

We now tackle the product of two binary variables $F_{i,f}$ and $F_{k,f}$ in 
\eqref{eq:P1_const3}. To this end, we linearize the product $F_{i,f}F_{k,f},~\forall i,~\forall k,~\forall f,$ by substituting it with a new slack variable $q_{i,k,f},~\forall i,~\forall k,~\forall f,$ along with the following three new constraints as
\begin{IEEEeqnarray*}{lcl}\label{eq8}
   &q_{i,k,f}\leq F_{i,f},  \IEEEyesnumber \IEEEyessubnumber* \label{eq8_a}\\
   &q_{i,k,f}\leq F_{k,f}, \label{eq8_b}\\
   &q_{i,k,f}\geq F_{i,f}+F_{k,f}-1. \label{eq8_c}
\end{IEEEeqnarray*}

Next, we linearize the products $F_{i,f}P_i$ and $q_{i,k,f}P_k$, which are appearing in the right hand side of \eqref{eq:obj_transform_a} and \eqref{eq:obj_transform_b}, respectively, as explained in \cite{ref11}. To this end, we substitute the product $F_{i,f}P_i,~\forall i,~\forall f,$ with a slack variable $t_{i,f},~\forall i,~\forall f,$ along with four new linear constraints, which ensure that the slack variable
equals the actual product, as
\begin{IEEEeqnarray*}{lcl}\label{eq9}
0\leq t_{i,f}\leq P_{\text{max}}F_{i,f},\IEEEyesnumber \IEEEyessubnumber* \label{eq9_a}\\
0\leq P_i-t_{i,f}\leq P_{\text{max}}(1-F_{i,f})\label{eq9_b},
\end{IEEEeqnarray*}
where $P_{\text{max}}=P_\text{t}^{\text{HAPS}}$, if UE $i$ is associated with the HAPS, otherwise $P_{\text{max}}=P_\text{t}^{\text{MBS}}$. Similarly, the product $q_{i,k,f}P_k,~\forall i,~\forall k,~\forall f,$ is substituted with $s_{i,k,f},~\forall i,~\forall k,~\forall f,$ along with four new linear constraints as 
\begin{IEEEeqnarray*}{lcl}\label{eq10}
&0\leq s_{i,k,f}\leq P_{\text{max}}q_{i,k,f},\IEEEyesnumber \IEEEyessubnumber* \label{eq10_a}\\
&0\leq P_k-s_{i,k,f}\leq P_{\text{max}}(1-q_{i,k,f})\label{eq10_b},
\end{IEEEeqnarray*}
where $q_{i,k,f}$ represents the product $F_{i,f}F_{k,f}$.
Next, we deal with the non-convex constraints by approximating them. Particularly, the left hand side of inequality \eqref{eq:obj_transform_a} is replaced with its upper bound convex approximation as \cite{ref9, ref10}
\begin{equation}{\label{eq11}}
z_i\beta_i\leq F(z_i,\beta_i,\xi_i) \triangleq \frac{\xi_i}{2}\beta_i^2+\frac{1}{2\xi_i}z_i^2;    ~\forall{\xi_i>0}.
\end{equation}

The equality holds when $\xi_i=z_i/\beta_i$. Replacing the left hand side of \eqref{eq:obj_transform_a} by the upper bound in \eqref{eq11}, and the products $F_{i,f}F_{k,f}$, $F_{i,f}P_i$ and $q_{i,k,f}P_k$ by $q_{i,k,f}$, $t_{i,f}$, and $s_{i,k,f}$, respectively, the approximated problem to be solved in the $n$th SCA iteration can be written as
\begin{IEEEeqnarray*}{lcl}\label{eq:P2}
&\underset{\mathbf{F},\mathbf{P},\mathbf{q},\mathbf{t},\mathbf{s},\mathbf{z}, \boldsymbol{\beta}}{\text{maximize}}\,\, & \min_{\forall i}\quad z_i \IEEEyesnumber \IEEEyessubnumber* \label{eq:P2_Obj}\\
&\text{s.t.} & \frac{\xi^{(n)}_i}{2}\beta_i^2+\frac{1}{2\xi^{(n)}_i}z_i^2\leq \tilde{P}_{\text{R}_i}, \, ~\forall i,\label{eq:P2_const1}\\
&& \beta_i\geq \tilde{I}_i+P_{\text{N}}, \,~\forall i, \label{eq:P2_const2}\\
&& \sum_{j=1}^{N_{\text{B}}+1}{\sum_{k=1,k\not=i}^{N_{\text{u}}}{A_{i,j}A_{k,j}\sum_{f=1}^{N_{\text{f}}}{q_{i,k,f}}}}=0, \, ~\forall i, \qquad \label{eq:P2_const3}\\
&& s_{i,k,f}\geq0, t_{i,f}\geq0, z_i\geq0, \, ~\forall i, ~\forall k, ~\forall f, \qquad \label{P2_const4}\\
&& q_{i,k,f} \in \mathbb{B}, \, ~\forall i, ~\forall k, ~\forall f, \label{eq:P2_const5}\\
&&  (\ref{eq:P1_const1}), ~(\ref{eq:P1_const2}), ~(\ref{eq:P1_const4}), ~(\ref{eq:P1_const5}), ~(\ref{eq:P1_const6}),~(\ref{eq8}), ~(\ref{eq9}), ~(\ref{eq10}),
\IEEEyessubnumber*
\end{IEEEeqnarray*}
where $\mathbf{q}, ~\mathbf{t}, ~\mathbf{s}, ~\mathbf{z}, ~\boldsymbol{\beta}$ are the matrices and vectors that collect variables $q_{i,k,f}, ~t_{i,f}, ~s_{i,k,f}, ~z_i, ~\beta_i,  ~\forall i,~\forall k,~\forall f$, respectively, and
\begin{equation}{\label{eq14}}
\tilde{P}_{\text{R}_i}=(1-A_{i,N_{\text{B}}+1})\sum_{f=1}^{N_{\text{f}}}{\frac{t_{i,f}\sum_{r=1}^{N_{\text{R}}}{\left|h_{r,i,\text{J}_i,f}\right|^2}}{\text{PL}_{i,\text{J}_i}}}\\
+A_{i,N_{\text{B}}+1}\frac{P_{i}G_{i}(\theta_i,\phi_i)}{\text{PL}_{i,N_{\text{B}}+1}},
\end{equation}
\begin{equation}{\label{eq15}}
\tilde{I}_i=\sum_{k=1,i\not=k}^{N_{\text{u}}}(1-A_{k,N_{\text{B}}+1})\\
\frac{\sum_{f=1}^{N_{\text{f}}}{s_{i,k,f}\sum_{r=1}^{N_{\text{R}}}{\left|h_{r,i,J_k,f}h^*_{r,k,J_k,f}\right|}}}{\text{PL}_{i,J_k}}\\
+A_{k,N_{\text{B}}+1}\frac{G_k(\theta_i,\phi_i)\sum_{f=1}^{N_{\text{f}}}{s_{i,k,f}}}{\text{PL}_{i,J_k}}.
\end{equation}

Due to the presence of binary variables, \eqref{eq:P2} is still a non-convex problem. To deal with the binary variables, we can use the Gurobi solver (version  9.5.2 build v9.5.2rc0) which is capable of solving binary-integer programming problems.

We are now in a position to present an algorithm that provides the optimal solution for the approximate problem \eqref{eq:P2} which is the sub-optimal solution for the original problem \eqref{eq:P1}. The pseudocode of the resource allocation design is outlined in Algorithm~\ref{alg:alg1}. According to the algorithm, \eqref{eq:P2} is solved iteratively until the convergence of the objective function value or maximum number of SCA iterations, $I_{\text{iter}}$, is reached, whichever first.
\begin{algorithm}[h!]
\caption{Proposed iterative resource allocation design algorithm for vHetNets.}\label{alg:alg1}
\begin{algorithmic}[1]
\STATE \textbf{Input:}~$N_{\text{f}}, ~N_{\text{B}}, ~N_{\text{u}}, ~N_{\text{R}}, ~N_{\text{SC}}, ~P_{\text{N}}, ~\mathbf{H}, ~\mathbf{G}(\theta,\phi), ~\mathbf{A}, ~P_\text{t}^{\text{HAPS}}, ~P_\text{t}^{\text{MBS}}, ~\mathbf{PL}, ~I_{\text{iter}}$.\\
\STATE \textbf{Output:} $\mathbf{F}^*, ~\mathbf{P}^*$.\\
\STATE \text{Initialize} $\boldsymbol{\xi}$ \text{and set} $n:=0.$\\
\STATE \text{Repeat until convergence or until $n \leq N:$}\\
\hspace{0.5cm} \text{- solve (\ref{eq:P2}) to find $\mathbf{F}^{(n)*}, ~\mathbf{P}^{(n)*}, ~\mathbf{q}^{(n)*}, ~\mathbf{t}^{(n)*}, ~\mathbf{s}^{(n)*}, ~\mathbf{z}^{(n)*}, ~\boldsymbol{\beta}^{(n)*}$.}\\
\hspace{0.5cm} \text{- update $\boldsymbol{\xi}^{(n+1)}:=\mathbf{z}^{(n)*}/\boldsymbol{\beta}^{(n)*}, ~n:=n+1$.}
\end{algorithmic}
\label{alg1}
\end{algorithm}

In Algorithm~\ref{alg1}, $\boldsymbol{\xi}$, $\mathbf{G}$, $\mathbf{PL}$, and $\mathbf{H}$, denote vector and matrices, which collect $\xi_i, ~\forall i$, $G_i(\theta,\phi), ~\forall i$, $\text{PL}_{i,j}$, $\forall i$, $\forall j$, and $h_{r,i,j,f}, ~\forall r, ~\forall i, ~\forall j, ~\forall f$, respectively. In each SCA iteration, the algorithm solves MILP consisting of a second-order cone program (SOCP)\cite{SOCP} and a mixed-integer program (MIP). The worst-case computational cost of the algorithm is mainly determined by the branch-and-bound method to solve MIP, and is given as $\mathcal{O}(2^{N_{\text{u}}N_{\text{f}}})$.

\section{Numerical Result and Discussion}\label{Sec:Results}
In this section, we numerically evaluate the performance of the proposed iterative Algorithm~\ref{alg1} and compare it with some baseline resource allocation schemes for vHetNet and standalone terrestrial network. In particular, we compare six scenarios (one proposed and five baselines); Scenario 1: joint power-subcarrier allocation (Algorithm~\ref{alg:alg1}) for vHetNet ($4$ MBSs and $1$ HAPS), Scenario 2: power control with random subcarrier allocation for vHetNet (in this scenario, Algorithm~\ref{alg:alg1} is implemented using fixed random $\mathbf{F}$), Scenario 3: optimized subcarrier allocation with no power control for vHetNet (in this scenario, in each tier, equal power is allocated and $\mathbf{F}$ is designed using Algorithm~\ref{alg:alg1}), Scenario 4: random subcarrier allocation with no power control for vHetNet (in this scenario, $\mathbf{F}$ is a fixed random matrix, and in each tier, equal power is allocated to UEs), Scenario 5: joint power-subcarrier allocation for terrestrial network ($4$ MBSs, no HAPS), and Scenario 6: joint power-subcarrier allocation for terrestrial network ($5$ MBSs, no HAPS).

Algorithm~\ref{alg:alg1} is implemented in Python and uses Gurobi \cite{ref13} as the solver. In simulations, we consider a circular urban area of radius $2$ km with $N_{\text{u}} = 16$ uniformly distributed UEs. Since the complexity of the algorithm is high due to binary variables, a small-scale network is considered to understand the behaviour of the proposed algorithm. For each scenario, the results are obtained based on $1,000$ i.i.d. network topologies. Table~\ref{tab:table1} summarizes the list of other parameters and their values used in algorithm 1 and simulations. The SE in this section is calculated as $\text{SE}=\log_2(1+\gamma)$.

\begin{table}[!t]
\caption{Simulation Parameters.\label{tab:table1}}
\centering
\begin{tabular}{|c||c|}
\hline
\textbf{Parameter} & \textbf{Value}\\
\hline
Center frequency ($f_c$) & $2.545$ GHz\\
\hline
$N_{\text{f}}$ & $4$\\
\hline
$N_{\text{R}}$ & $8$\\
\hline
$N_{\text{SC}}$ & $1$\\
\hline
$P_t^{\text{HAPS}}, P_t^{\text{MBS}}$ & $55$ dBm \cite{ref14}, $43$ dBm\\
\hline
$P_{\text{N}}$ & $-105$ dBm\\
\hline
HAPS altitude & $20$ Km\\
\hline
$N_{\text{V}} \times N_{\text{H}}$ & $64\times64$\\
\hline
HAPS $3$dB beamwidth & $65$ degrees\cite{ref14}\\
\hline
HAPS antenna element gain & $8$ dBi\cite{ref14}\\
\hline
HAPS antenna element front to back ratio & $30$ dB\cite{ref14}\\
\hline
Maximum number of SCA iterations ($I_{\text{iter}}$) & $20$\\
\hline
\end{tabular}
\end{table}
\begin{figure*}[!t]
\centering
\captionsetup{justification=centering}
\includegraphics[width=0.5\linewidth]{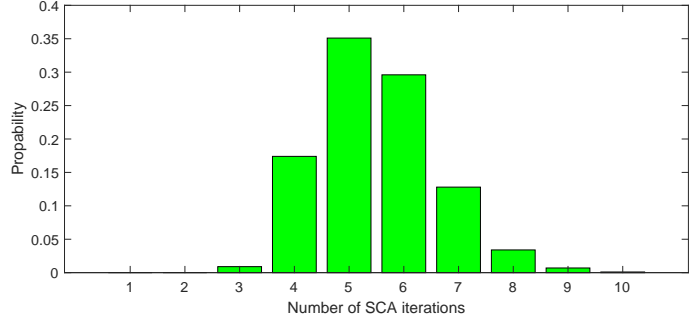}\label{Fig2}
\vspace{-0.2cm}
\caption{\small Convergence behaviour of the proposed Algorithm~\ref{alg:alg1}.}\label{fig_2}
\end{figure*}
First, we present the convergence behaviour of the proposed algorithm. Fig.~\ref{fig_2} depicts the probability distribution of the number of SCA iterations. It can be observed that, for all 1,000 topologies, the algorithm converges mostly in five and six iterations and at most in ten iterations.
\begin{figure*}[t]
    \centering
    \captionsetup{justification=centering}
    \subfloat[CDF of minimum spectral efficiency.]{\includegraphics[width=0.49\columnwidth,height=4.8cm]{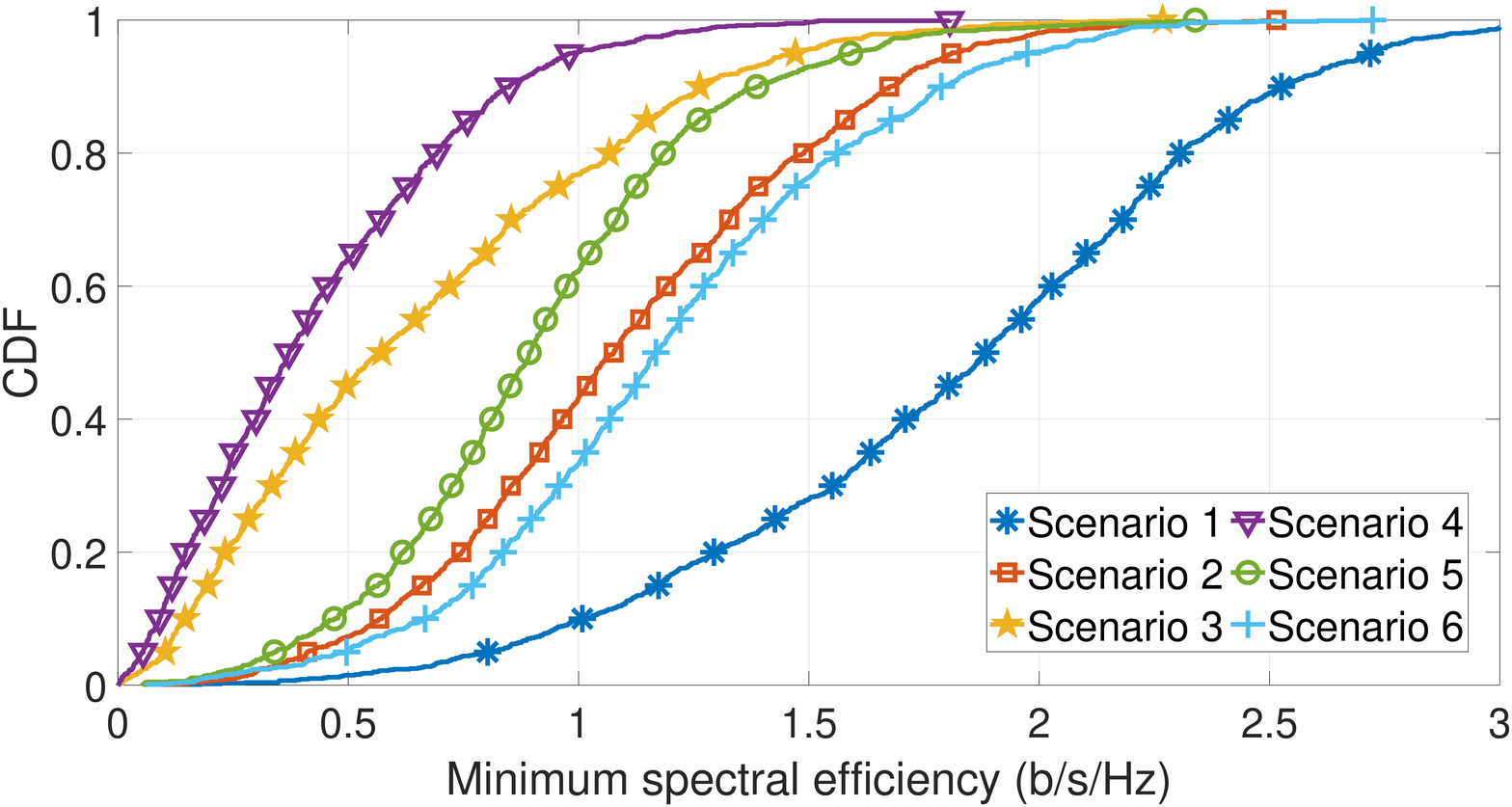}\label{fig_3:a}}
    \subfloat[CDF of spectral efficiency.]{\includegraphics[width=0.49\columnwidth,height=4.8cm]{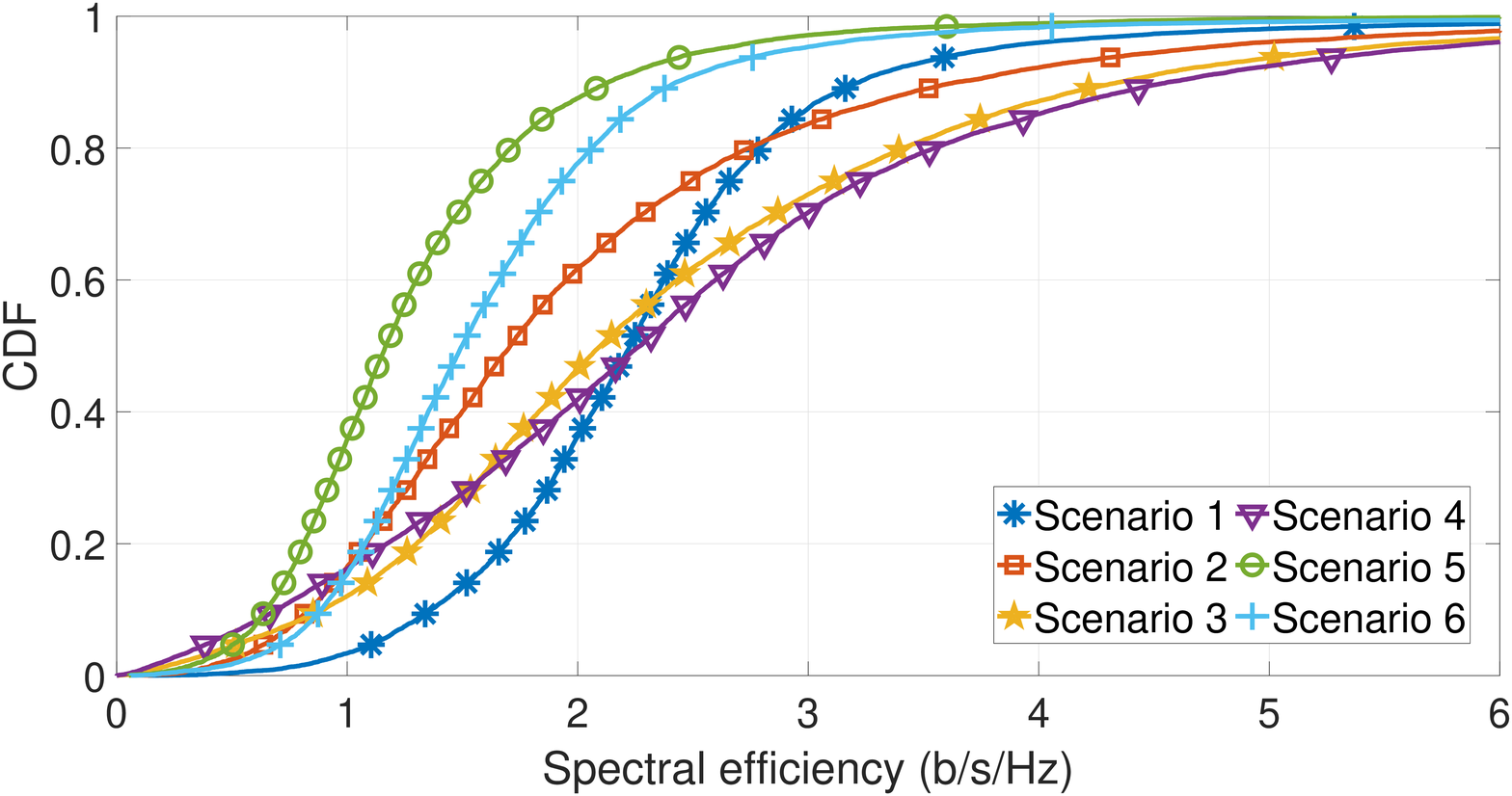}\label{fig_3:b}}
    \vspace{-0.2cm}
    \caption{\small Statistical behavior of six scenarios. Scenario~1:~vHetNet~(4~MBSs~+~1~HAPS)~with~jointly~optimized~power-subcarrier~allocation, Scenario 2: vHetNet (4 MBSs + 1 HAPS) with random subcarrier~allocation~and~optimized~power~control,~Scenario~3:~vHetNet~(4~MBSs~+ 1 HAPS) with optimized subcarrier allocation and no power control, Scenario~4:~vHetNet~(4~MBSs~+~1~HAPS)~with~random~subcarrier allocation and no power control, Scenario 5: Terrestrial network (4 MBSs, no HAPS)~with~jointly~optimized~power-subcarrier~allocation, and Scenario 6: Terrestrial network (5 MBSs, no HAPS) with jointly optimized power-subcarrier allocation.}
    \label{fig_3}
\end{figure*}

Fig.~\ref{fig_3} investigates the statistical behaviour of six aforementioned scenarios. Particularly, Fig.~\ref{fig_3:a} plots the CDFs of the minimum SE for six scenarios. From the figure, we can conclude three important observations. Firstly, the proposed joint power-subcarrier allocation design obtained using Algorithm~\ref{alg:alg1}, in vHetNet (i.e., Scenario 1) provides the best SE to the worst UE as compared to those obtained through the rest of the scenarios. Secondly, by comparing Scenarios 1, 5, and 6, it can be observed that using HAPS as the complementary platform to the terrestrial network (i.e., Scenario 1) improves the worst UE SE performance significantly over the standalone terrestrial network (i.e., Scenario 5). Even deploying one additional MBS to the terrestrial network (i.e., Scenario 6) is not able to achieve the performance of Scenario 1. Lastly, it can be observed that transmit power control plays an important role in mitigating interference in vHetNets. The performance of the scenarios without power control (i.e., Scenarios 3 and 4) is even worse than the standalone terrestrial network (i.e., Scenarios 5 and 6). Moreover, compared with Scenario 4, it can also be observed from Scenarios 2 and 3 that the minimum SE of UEs is more sensitive to power control than subcarrier optimization in vHetNets.

Fig.~\ref{fig_3:b} plots the CDFs of SE for six scenarios. It can be observed that the value of the fifth percentile for Scenario 1 is the highest. This observation corroborates with the observation, from Fig.~\ref{fig_3:a} that the proposed Algorithm~\ref{alg:alg1} in vHetNet improves the SE of worst UEs compared to other scenarios. Moreover, vHetNet with power control (i.e., Scenarios 1 and 2) provides better performance for worst UEs compared to Scenarios 3 and 4, where no power control is employed. The reason for this behaviour is the high interference experienced by the UEs associated with MBSs due to the high transmit power of HAPS in Scenarios 3 and 4. Moreover, we observe that standalone terrestrial networks (i.e., Scenarios 5 and 6) have a lower variance for SE compared to vHetNets (i.e., Scenarios 1, 2, 3, and 4).

\section{Conclusion}\label{Sec:conclution}
We studied an important problem of interference in an integrated HAPS-terrestrial network (vHetNet). To this end, we proposed an optimum power-subcarrier allocation scheme and formulated a MMF optimization problem accordingly. Since the original problem is a MINLP and challenging to solve, we reformulated the problem to make it tractable and convex. Then, we developed a SCA-based iterative algorithm that solves the reformulated problem. Simulation results showed that vHetNets with optimized power-subcarrier allocation offer higher SE to worst UEs than standalone terrestrial network scenarios. This observation proves the necessity of a complementary platform for terrestrial networks to improve the performance. Moreover, in vHetNets, interference is more sensitive to transmit power control than subcarrier optimization.

\vfill

\end{document}